
\magnification=1200
\baselineskip=12pt
\tolerance=100000
\overfullrule=0pt
\def\dbar {\hbox{D\kern-0.52em\raise+0.45ex\hbox{--}\kern+0.2em}}
\rightline{UR-1320\ \ \ \ \ \ }
\rightline{ER40685-770}

\baselineskip=20pt
\centerline{\bf MORE ON SYMMETRIES IN HEAVY QUARK EFFECTIVE THEORY}

\vskip 2cm

\centerline{Ashok Das}
\centerline{and}
\centerline{V. S. Mathur}
\centerline{Department of Physics and Astronomy}
\centerline{University of Rochester}
\centerline{Rochester, NY 14627}
\vskip 1cm

\centerline{\underbar{\bf Abstract}}

We present a general classification of all normal and ``chiral" symmetries
 of heavy quark effective theories.  Some peculiarities and conondrums
associated with the ``chiral" symmetries are discussed.

\vfill\eject

\noindent {\bf I. Introduction:}
\smallskip
The physics of processes involving hadrons containing a heavy quark
can be described adequately by the heavy quark effective theory (HQET).
HQET [1-5] is a simple theory and has many expected symmetries such as the spin
and flavor symmetries.  These are analogous to the spin and flavor
symmetries which one would expect in quantum electrodynamics (QED) in the
infrared limit [6].  However, in two recent interesting papers [7],
 it was shown
that the lowest order HQET contains extra unexpected symmetries of
 the ``chiral" type.  These are kind of unexpected symmetries which
have been argued [7] to be spontaneously broken.  In this
paper, we study these symmetries more systematically and bring out some
peculiarities associated with them.

The organization of the paper is as follows.  It is known that the heavy
quark theory can be obtained from Quantum
 Chromodynamics (QCD) in the nonrelativistic limit through
a Foldy-Wouthuysen transformation [8-10].
  In sec. II, we briefly discuss the
Foldy-Wouthuysen transformations and the series expansion of HQET in
inverse powers of the heavy quark mass.  In sec. III, we classify all the
normal and ``chiral" symmetries of HQET.  In sec. IV we bring out various
peculiarities associated with the ``chiral" symmetries
 with a brief conclusion in sec. V.
\medskip

\noindent {\bf II. Effective Nonrelativistic Theory:}
\smallskip

Let us consider a massive, free fermion theory described by
$${\cal L} = \overline \psi \left(i \gamma^\mu \partial_\mu - m \right)
 \psi \qquad\qquad \qquad \mu = 0, 1, 2, 3 \eqno(1)$$
We use the metric $\eta^{\mu \nu} = (+,-,-,-)$ and our Dirac matrices have
the representation
$$\gamma^0 = \pmatrix{I &0\cr
\noalign{\vskip 5pt}%
0 &-I\cr}
\qquad \qquad \qquad
\gamma^i = \pmatrix{0 &\sigma_i\cr
\noalign{\vskip 5pt}%
-\sigma_i &0\cr}$$
\rightline{(2)}
$$\gamma_5 = i \gamma^0 \gamma^1 \gamma^2 \gamma^3 =
\pmatrix{0 &I \cr
\noalign{\vskip 5pt}%
I &0\cr}$$
where the $\sigma_i$'s $(i = 1,2,3)$ represent the three Pauli matrices.
Note that the matrices $\gamma^i$ couple the upper and lower two component
spinors of the four component $\psi$ while $\gamma^0$ does not.  In trying
to obtain the nonrelativistic limit, the goal is to decouple the
 upper and lower two
component spinors since a nonrelativistic fermion has only two degrees of
freedom.  This can be achieved through the Foldy-Wouthuysen
transformations [8].  In fact, it is straightforward to see that under the
redefinition of variables
$$\eqalign{\psi &\longrightarrow e^{{i \over 2m}\ \vec \gamma \cdot
\vec \nabla \alpha ({| \vec \nabla | \over m})} \psi\cr
\overline \psi &\longrightarrow \overline \psi \
e^{- {i \over 2m}\ \vec \gamma \cdot {\buildrel \leftarrow \over \nabla}
 \alpha ({| {\buildrel \leftarrow \over \nabla}
| \over m})} \cr}\eqno(3)$$
where the gradients in the $\overline \psi$ redefinition act on
$\overline \psi$ and
$$\alpha \left( {| \vec \nabla | \over m}\right) = {m \over | \vec
\nabla |}\ \tanh^{-1} \left( {| \vec
\nabla | \over m}\right) \eqno(4)$$
the Lagrangian of Eq. (1) takes the form
(we neglect surface terms throughout the paper)
$${\cal L} = \overline \psi i \gamma^0 \partial_0 \psi - \overline \psi
\left( m^2 - \vec \nabla^2 \right)^{1/2} \psi \eqno(5)$$
The upper and lower two component spinors are now decoupled and the
nonrelativistic limit can be obtained by expanding $(m^2 - \vec
\nabla^2)^{1/2}$ in inverse powers of mass.  We note here that the second
term in Eq. (5) can be removed through the redefinition
$$\eqalign{\psi &= e^{-i (m^2 - \vec \nabla^2)^{1/2}
\gamma^0 t} \psi^\prime\cr
\overline \psi &= \overline \psi^\prime e^{i (m^2 -
 {\buildrel \leftarrow \over \nabla}^2)^{1/2}
\gamma^0 t} \cr}\eqno(6)$$
where once again the derivatives in the second line are supposed to act on
$\overline \psi^\prime$.  The redefinitions in Eq. (6) merely correspond to
the time evolution of the four component spinors and consequently, the
spinors $\psi^\prime$ have no time dependence.  This can also be seen from
the fact that the Lagrangian, in terms of $\psi^\prime$, has the static
form
$$\eqalign{{\cal L} &= \overline \psi i \gamma^0 \partial_0 \psi -
\overline \psi (m^2 - \vec \nabla^2)^{1/2} \psi\cr
&= \overline \psi^\prime i \gamma^0 \partial_0 \psi^\prime\cr}\eqno(7)$$

The Foldy-Wouthuysen transformations are highly nonlocal and
have a closed form only for the free fermion theory.  In the presence of
interactions, the Foldy-Wouthuysen transformations can be carried out order
by order to any given order in ${1 \over m}$ in the following way [8-10].
  Let us
consider a massive fermion interacting minimally with a gauge field (The
structure of the gauge group is irrelevant for our discussion.)  described
by the Lagrangian
$${\cal L} = \overline \psi \left( i \gamma^\mu D_\mu - m \right)
\psi \eqno(8)$$
where
$$D_\mu \psi = \left( \partial_\mu + A_\mu \right) \psi \eqno(9)$$
with $A_\mu$ belonging to the appropriate representation of the fermions in
the case of a non Abelian symmetry group.  If we now redefine variables as
(analogous to Eq. (3) with $\alpha = 1$)
$$\eqalign{\psi &\longrightarrow e^{i \vec \gamma \cdot \vec D/2m} \psi\cr
\overline \psi &\longrightarrow \overline \psi e^{-i \vec \gamma
\cdot {\buildrel \leftarrow \over D} /2m} \cr}\eqno(10)$$
then the Lagrangian in the new variables takes the form
$${\cal L} = \overline \psi \left( i \gamma^0 D_0 - m \right)
\psi + \sum^\infty_{n=1} {1 \over m^n}\
\overline \psi O_n \psi \eqno(11)$$
where
$$\eqalign{O_n = {1 \over n!} \big( - {i \over 2} \big)^n i \gamma^0
\big[ \vec \gamma \cdot \vec D ,
 \big[ \vec \gamma \cdot \vec D, \dots
   \big[ \vec \gamma \cdot
\vec D, D_0 \big]
 &\ \dots \dots\  \big] \cr
\noalign{\vskip -10pt}
&\leftarrow n \rightarrow\cr}$$
\smallskip
$$ + {n \over (n + 1)!} \ \big( i \vec \gamma \cdot \vec D
\big)^{n+1} \qquad\qquad\qquad n \ge 1 \eqno(12)$$
In particular, we note that
$$O_1 = {1 \over 2}\ \vec D^2 - {1 \over 4}\ \gamma^\mu \gamma^\nu
F_{\mu \nu} \eqno(13)$$
where the field strength is defined to be
$$F_{\mu \nu} = \left[ D_\mu , D_\nu \right]$$

It is worth noting here that while $\gamma^0$ is diagonal, the matrices $O_n$
 are not in general
 block diagonal and, therefore, would couple the upper and the
lower two component spinors.  However, order by order, they can be
block diagonalized through appropriate field redefinition.  Thus, for example,
let us assume that the matrices are
 block diagonal up to order $k$ and that the
first nondiagonal matrix is $O_{k+1}$.  This can be uniquely separated into
a diagonal and an off-diagonal part as
$$O_{k+1} = O^C_{k+1} + O^A_{k+1} \eqno(14)$$
where the diagonal matrix $O^C_{k+1}$ can be identified with
$$O^C_{k+1} = {1 \over 2} \left( O_{k+1} + \gamma^0 O_{k+1} \gamma^0
\right)\eqno(15)$$
while the off-diagonal matrix $O^A_{k+1}$ has the form
$$O^A_{k+1} = {1 \over 2} \left( O_{k+1} - \gamma^0 O_{k+1}
\gamma^0 \right) \eqno(16)$$
It follows now that
$$\eqalign{\big[ \gamma^0 , O^C_{k+1} \big] &= 0 \cr
\big[ \gamma^0 , O^A_{k+1} \big]_+ &= 0\cr}\eqno(17)$$
It is now straightforward to see that under a field redefinition
($\gamma^0 O^\dagger_n \gamma^0 = O_n$ for hermiticity of the
 Lagrangian in Eq. (11))
$$\eqalign{\psi &\longrightarrow e^{O^A_{k+1} / 2m^{k+2}} \psi\cr
\overline \psi &\longrightarrow \overline \psi e^{O^A_{k+1} /2m^{k+2}}\cr}
\eqno(18)$$
the Lagrangian takes the form
$${\cal L}
 = \overline \psi \left( i \gamma^0 D_0 - m \right) \psi
+ \sum^\infty_{n=1} {1 \over m^n}\
\overline \psi \widetilde O_n \psi \eqno(19)$$
where
$$\eqalign{\widetilde O_n &= O_n \quad {\rm for} \quad
n \leq k\cr
&= O_{k+1} - O^A_{k+1} = O^C_{k+1} \quad {\rm for}\quad n = k+1 \cr}
\eqno(20)$$
The higher order matrices (for $n > k+1$) are more complicated, but the
philosophy is clear.  Namely, order by order one can
 block diagonalize the
matrices $O_n$ through a series of Foldy-Wouthuysen transformations.  Once
the diagonalization is carried out, the Lagrangian will have the form
$${\cal L} = \overline \psi \left( i \gamma^0 D_0 - m \right) \psi +
\sum^\infty_{n=1} {1 \over m^n}\ \overline \psi O_n \psi \eqno(21)$$
where all the $O_n$ matrices will be
 block diagonal.  This, then, would represent
the nonrelativistic limit of the full theory and has a power series
expansion in the inverse power of the heavy quark mass.  We also note here
that the mass term can be transformed away through a phase
 redefinition of the field of
the form
$$\eqalign{\psi &= e^{- i m \gamma^0 t} \psi^\prime\cr
\overline \psi &= \overline \psi^\prime
e^{i m \gamma^0 t} \cr} \eqno(22)$$
so that the Lagrangian in terms of the $\psi^\prime$ variable becomes
$${\cal L} = \overline \psi^\prime i \gamma^0 D_0 \psi^\prime +
\sum^\infty_{n=1} {1 \over m^n}\ \overline \psi^\prime
O_n \psi^\prime \eqno(23)$$
Unlike the free fermion theory, however, the transformation of Eq. (22)
does not represent the complete time evolution of the fermions and
consequently, $\psi^\prime$ carries time dependence.  Furthermore, we note
that the entire discussion can be cast in a manifestly Lorentz covariant
form by introducing a velocity four vector $v^\mu$ which satisfies [4,10]
$$v^\mu v_\mu = 1 \eqno(24)$$
and which allows us to replace
$$\eqalign{\gamma^0 D_0 &\longrightarrow  \rlap\slash{v} v \cdot D\cr
\vec \gamma \cdot \vec D &\longrightarrow  \rlap \slash{\rm D}
- \rlap\slash{v}  v \cdot D\cr}\eqno(25)$$
In the special frame where $v^\mu = (1,0,0,0)$,
we obtain
Eq. (21) or (23) which is the form of the Lagrangian we will use in our
discussion for simplicity.

\medskip
\vfill\eject
\noindent {\bf III.  Classification of Symmetries:}

\smallskip

Let us consider the zeroth order nonrelativistic Lagrangian
 of Eq. (21) or (23), namely,
$$\eqalign{{\cal L}_0 &= \overline \psi \big( i \gamma^0
D_0 - m \big) \psi\cr
&= \overline \psi^\prime i \gamma^0 D_0 \psi^\prime \cr}\eqno(26)$$
The nongauge symmetries of this Lagrangian are now straightforward to
classify particularly in terms of the $\psi^\prime$ variables.    Let us
consider the transformations
$$\eqalign{\psi^\prime &\longrightarrow e^{iA} \psi^\prime\cr
\overline \psi^\prime &\longrightarrow \overline \psi^\prime
\gamma^0 e^{- i A^\dagger} \gamma^0\cr}\eqno(27)$$
where $A$ is a space-time independent 4 $\times$ 4 matrix.

\noindent a) Normal Symmetries:

It is clear that when
$$\gamma^0 A^\dagger \gamma^0 = A\eqno(28)$$
and
$$[ \gamma^0 , A ] = 0 \eqno(28^\prime)$$
the transformations in Eq. (27) will be a symmetry of the
Lagrangian.  In this case, the matrices $A$
 will be block diagonal and hence will not mix the upper and lower spinor
functions.  There are eight such linearly independent 4 $\times$ 4 matrices
and they are
$$A = 1,\ \gamma^0,\ - \gamma_5 \gamma^i,\ \sigma^{ij} =
{i \over 2}\  \big[ \gamma^i , \gamma^j \big] \eqno(29)$$
It is interesting to note that these eight matrices can be grouped and
rewritten as
$$\eqalign{K^\mu_N &= \pmatrix{\sigma^\mu &0\cr
\noalign{\vskip 5pt}%
0 & \sigma^\mu\cr}\cr
\noalign{\vskip 4pt}%
M^\mu_N &= \pmatrix{\sigma^\mu &0\cr
\noalign{\vskip 5pt}%
0 &- \sigma^\mu\cr}\cr}\eqno(30)$$
where $\sigma^\mu = ({\bf 1}, \vec \sigma)$ (namely, the identity and the
Pauli matrices) denote the four linearly independent $2 \times 2$ matrices
which can act on a two dimensional spinor space.

\vfill\eject
\noindent b) ``Chiral" Symmetries:

On the other hand, if
$$\gamma^0 A^\dagger \gamma^0 = - A \eqno(31)$$
and
$$\big[ \gamma^0 , A \big]_+ = 0\eqno (31^\prime)$$
then the transformations in Eq. (27) will also be a symmetry of the
Lagrangian.  In this case, the matrices $A$ will be off-diagonal and hence
will necessarily
 mix the upper and lower spinor functions.  There are eight such
independent 4 $\times$ 4 matrices and they are
$$A = \gamma_5,\ i \gamma_5 \gamma^0,\ i \gamma^i, \ \gamma^0
\gamma^i \eqno(32)$$
It is easy to see that these eight matrices  can also be grouped and
rewritten as
$$\eqalign{K^\mu_C &= \pmatrix{0 &\sigma^\mu\cr
\noalign{\vskip 5pt}%
\sigma^\mu &0\cr}\cr
\noalign{\vskip 4pt}%
M^\mu_C &= \pmatrix{0 &i \sigma^\mu\cr
\noalign{\vskip 5pt}%
- i \sigma^\mu &0\cr}\cr}\eqno(33)$$
Furthermore, they can be related to the generators of the normal symmetries
as
$$\eqalign{K^\mu_C &= \gamma_5 K^\mu_N\cr
M^\mu_C &= - i \gamma_5 M^\mu_N\cr}\eqno(34)$$

Basically, therefore, the sixteen generators of the Clifford algebra split
into the generators of the two classes of symmetries (of course, with
appropriate normalization) depending on whether they commute with $\gamma^0
$ or anticommute with it.  (In the covariant language it is the
commutation or anticommutation with $\rlap\slash{v} = \gamma^\mu v_\mu$
which determines the two classes of symmetries.)
It is also worth noting here that even though we have shown these
transformations to be symmetries of the zeroth order Lagrangian in Eq.
(26), it is quite straightforward to see that these are symmetries of the full
free fermion Lagrangian in Eq. (7).  In contrast, when interactions are
present even the first order correction in the effective Lagrangian
violates all the symmetries except the normal symmetries generated by
$$A = {\bf 1},\ \gamma^0 \eqno(35)$$
which hold order by order
 to all orders in ${1 \over m}$.

\medskip

\noindent {\bf IV. Peculiarities of ``Chiral" Symmetries:}

\smallskip

The ``chiral" symmetries of the heavy quark effective theory are quite
unusual and in this section, we will try to bring out some of the
peculiarities of such symmetries using $\gamma_5$ symmetry as an example.
The discussion holds for all the other ``chiral" symmetries as well.  Let
us consider for simplicity the zero momentum limit of the free fermion
theory.  The Lagrangian in this case takes the form (see Eq. (7))
$$\eqalign{{\cal L} &= \overline \psi (i \gamma^0 \partial_0 - m ) \psi \cr
&= \overline \psi^\prime i \gamma^0 \partial_0 \psi^\prime\cr}\eqno(36)$$
with
$$\psi = e^{-i m \gamma^0 t} \psi^\prime \eqno(37)$$
Under the $\gamma_5$-transformations
$$\eqalign{\psi^\prime &\longrightarrow e^{i \epsilon \gamma_5}
\psi^\prime\cr
\overline \psi^\prime &\longrightarrow \overline \psi^\prime e^{i \epsilon
\gamma_5} \cr}\eqno(38)$$
the Lagrangian in Eq. (36)
 is invariant.  As we have discussed earlier, $\gamma_5$ is
an off-diagonal matrix and consequently, this transformation mixes the
upper and the lower two component spinors.  In the second quantized
language, this would correspond to mixing of particles and antiparticles.  In
fact, if we expand the field variables as usual as
$$\psi = \sum^2_{s=1} e^{-imt} a(s) u(s) + e^{imt}
b^\dagger (s) v(s) \eqno(39)$$
or equivalently
$$\psi^\prime = \sum^2_{s=1} a(s) u(s) + b^\dagger (s) v(s) \eqno(40)$$
where
$$u(1) = \pmatrix{1\cr
\noalign{\vskip 5pt}%
0\cr
\noalign{\vskip 5pt}%
0\cr
\noalign{\vskip 5pt}%
0\cr} \qquad u(2) = \pmatrix{0\cr
\noalign{\vskip 5pt}%
1\cr
\noalign{\vskip 5pt}%
0\cr
\noalign{\vskip 5pt}%
0\cr}
\qquad
v(1) = \pmatrix{0\cr
\noalign{\vskip 5pt}%
0\cr
\noalign{\vskip 5pt}%
1\cr
\noalign{\vskip 5pt}%
0\cr}
\qquad
v(2) = \pmatrix{0\cr
\noalign{\vskip 5pt}%
0\cr
\noalign{\vskip 5pt}%
0\cr
\noalign{\vskip 5pt}%
1\cr}
\eqno(41)$$
then, it is straightforward to see that the generator of the
transformations in Eq. (38) takes the form
$$Q_5 = \sum^2_{s=1} (a(s)b(s) +
b^\dagger (s) a^\dagger (s)) \eqno(42)$$
This is the generator of a Bogoliubov transformation and as is well known, in
a quantum field theory, it leads to unitarily inequivalent Hilbert spaces
resulting in a spontaneous breakdown of the symmetry [11].

To better understand the meaning and the properties of these
 ``chiral" symmetries, we will carry out our discussion in the first
quantized  language for simplicity.  In this language, the equation of
motion for $\psi^\prime$ is given by
$$\eqalign{&i \gamma^0 \partial_0 \psi^\prime = 0\cr
&{\rm or},\ \ i\partial_0 \psi^\prime = 0 \cr}\eqno(43)$$
This simply shows that the static wave function can be any four-component
constant spinor.  The normal symmetries mix up the spinor components
preserving the probability as well as the Lorentz invariant normalization.
The ``chiral" symmetries, on the other hand, preserve the probability
associated with a given wave function but not $\overline \psi^\prime
\psi^\prime$.  This argument goes through even in the manifestly covariant
description where the equation of motion is given by
$$\eqalign{&i \rlap\slash{v} v \cdot \partial \psi^\prime = 0\cr
&{\rm or},\ \ i v \cdot \partial \psi^\prime = 0\cr
&{\rm or},\ \ i \partial_0 \psi^\prime = - i\  {\vec v \cdot
 \vec \nabla \over v^0}\ \psi^\prime \cr}\eqno(44)$$
The solutions, in this case, will correspond to
$$\psi^\prime = e^{-i \omega t + i \vec k \cdot \vec x} \chi
\eqno(45)$$
with
$$\omega = {\vec v \cdot \vec k \over v^0} \eqno(46)$$
and $\chi$ represents any space-time independent four component spinor.  It
is, interesting to note that the Hamiltonian for the $\psi^\prime$ system
is invariant under the normal as well as ``chiral" transformations.
(In fact, the Hamiltonian
 corresponding to the Lagrangian in
 Eq. (36) is trivially invariant since it
vanishes.)

Returning now to the original variables, $\psi$, we note that the
Lagrangian is invariant under the transformation
 ($\gamma_5$ as an example of the
``chiral" symmetries)
$$\eqalign{\psi \longrightarrow \widetilde \psi &= e^{-im \gamma^0 t}
e^{i \epsilon \gamma_5}
e^{im \gamma^0 t} \psi\cr
&= \big( \cos \epsilon + i \sin \epsilon\  \gamma_5 e^{2 im \gamma^0 t}\big)
\psi\cr}\eqno(47)$$
This is a time dependent transformation which nevertheless leaves the
Lagrangian invariant.  In fact, it consists of a time translation followed
by a $\gamma_5$-rotation and an inverse time translation.
We note that
$$\eqalign{\bigg( i \gamma^0\  {d \over dt} - m \bigg) \widetilde \psi &=
-i \sin \epsilon\  \gamma_5 (i \gamma^0 \cdot
2 im \gamma^0) e^{2im \gamma^0 t} \psi\cr
&\qquad + (\cos \epsilon - i \sin \epsilon\  \gamma_5 e^{2im\gamma^0 t})
i \gamma^0\  {\partial \psi \over \partial t}\cr
&\qquad - m (\cos \epsilon + i \sin \epsilon\  \gamma_5
e^{2 im \gamma^0 t} ) \psi\cr
&= (\cos \epsilon - i \sin \epsilon\  \gamma_5
e^{2 im \gamma^0 t} )
\bigg( i \gamma^0 \ {\partial \over \partial t}
- m \bigg) \psi \cr}\eqno(48)$$
In other words, if $\psi$ represents a solution of the Dirac equation, then
 so does
 $\widetilde \psi$.
Namely, the transformation takes a solution of the
dynamical equation to another solution.
 It is straightforward to see that under the
transformation of Eq. (47), a positive energy solution goes to a general
linear superposition of positive and negative energy solutions preserving
the probability.  Namely,
$$\eqalign{\psi (t) &= e^{-imt}u(s)\cr
& \longrightarrow \widetilde \psi (t) =
\cos \epsilon\  e^{-imt}u(s)
+ i \sin \epsilon\  e^{imt} v(s)\cr} \eqno(49)$$
where $u(s)$ and $v(s)$ are defined in Eq. (41).  Similarly, a negative
energy solution transforms to a general linear superposition of positive
and negative energy solutions conserving probability, namely,
$$\eqalign{\psi (t) &= e^{imt}v(s)\cr
& \longrightarrow \widetilde \psi (t) =
i \sin \epsilon\  e^{-imt}u(s)
+ \cos \epsilon\  e^{imt} v(s)\cr} \eqno(50)$$

It is slightly puzzling to note that under the symmetry transformation, an
eigenstate of energy ceases to be an energy eigenstate.
In this connection, we note
that for the system
 described in terms of the $\psi$ variables, the Hamiltonian is given by
$$H = m \gamma^0 \eqno(51)$$
This Hamiltonian is not invariant under the transformation
of Eq. (47) simply because
$\gamma^0$ does not commute with $\gamma_5$.  (The normal symmetries, on
the other hand, leave the Hamiltonian invariant.)  In other words, the
generator of the first quantized symmetry in Eq. (47), namely,
$$q_5 = \gamma_5 e^{2 im \gamma^0 t} \eqno(52)$$
does not commute with the Hamiltonian.  (This discussion can be carried out
equally well in the second quantized language.)  In fact, note that
$$\left[ q_5 , H \right] = 2 m \gamma_5 \gamma^0 e^{2 im
\gamma^0 t} \not= 0 \eqno(53)$$
The generator $q_5$, nevertheless, is conserved simply because it carries
explicit time-dependence.  Thus (with $\hbar = 1$)
$${d q_5 \over dt} = {\partial q_5 \over \partial t} + {1 \over i}
\ \left[ q_5 , H \right] = 0 \eqno(54)$$
(We note here that the generators of the normal symmetries, on the other
 hand, are time independent and commute with the Hamiltonian.)
This is quite counterintuitive to our general understanding of symmetries
where the generator of a symmetry commutes with the Hamiltonian of the
system [12].
A symmetry generator commuting with the Hamiltonian, of course, has
simultaneous energy eigenstates and states degenerate in energy are further
labelled by the quantum numbers of the conserved charge.  In contrast, in
the present case, the generator of symmetry does not commute with the
Hamiltonian and consequently, the energy eigenstates are not eigenstates of
the symmetry generator and cease to remain eigenstates of energy under a
symmetry transformation as is clear from Eqs. (49) and (50).

If we choose to work with the energy eigenbasis as is customary in
quantizing the theory, it is clear that the ``chiral" symmetries will no
longer appear to hold.  It is also not clear, whether the symmetry -- if it
is broken -- will be spontaneously broken as the analysis in the
$\psi^\prime$ variable seems to suggest.  The question of spontaneous
symmetry breaking in the case when a generator does not commute with the
Hamiltonian is not at all clear.  In fact, recall that conventionally
when a conserved charge commutes with the Hamiltonian, they have
simultaneous eigenstates and if the charge fails to annihilate the vacuum,
the symmetry is said to be spontaneously broken.  In contrast, in the present
case the energy eigenstates are not eigenstates of the symmetry generator
and consequently, nonannihilation of the vacuum by the symmetry generator
would appear inconsequential.  This is a puzzling feature.  However, we hasten
to point out here that there is one conventional symmetry which is quite
analogous to these ``chiral" symmetries.  We know that the Lorentz
boost generators, $K^i$, are explicitly time dependent.  From the Lorentz
algebra, we have
$$\left[ K^i , H \right] \sim P^i \eqno(55)$$
Yet, the boost generators are conserved, namely,
$${dK^i \over dt} = {\partial K^i \over \partial t} +
 {1 \over i}\ \left[ K^i , H \right] = 0 \eqno(56)$$
Choosing the energy eigenstates to quantize the theory would again seem to
imply that the Lorentz boost symmetry will be broken in such a case.
However, as we know from the studies of quantum field theory,  Lorentz
invariance holds true in spite of quantizing in the energy eigenbasis.
Drawing from this analogy, it is then tempting  to say that the ``chiral"
symmetries of the zeroth order heavy quark effective theory similarly would
not be violated even when quantized in the energy eigenbasis.  The analysis
in terms of the $\psi^\prime$ variable would then be a puzzle.
(In the $\psi^\prime$ formalism the Hamiltonian is invariant under the
transformation but the generator (see e.g. Eq. (42)) does not annihilate
the vacuum.)

\medskip

\noindent{\bf V. Conclusion:}

\smallskip

We have systematically classified all the symmetries of the heavy quark
effective theories.  We have shown that the properties of the ``chiral"
symmetries are quite different from our general understanding of symmetries
and we have tried to bring out the peculiarities of these ``chiral"
symmetries in a coherent manner.

\medskip

\noindent {\bf Acknowledgement}

\smallskip

One of us (A.D.)
 would like to thank Prof. S. Okubo for comments and Dr. R. Tzani
for discussing their work prior to publication.  This work was supported in
part by U.S. Department of Energy Grant No. DE-FG-02-91ER40685.

\vfill\eject

\noindent {\bf References}

\item{1.} W. E. Caswell and G. P. Lepage, Phys. Lett. {\bf B167} (1986)
437; G. P. Lepage and

\item{  } B. A. Thacker, Nucl. Phys. {\bf B4} (1988) 199.

\item{2.} N. Isgur and M. B. Wise, Phys. Lett. {\bf B232} (1989) 113; ibid
{\bf B237} (1990) 527.

\item{3.} E. Eichten and B. Hill, Phys. Lett. {\bf B234} (1990)
511; M. E. Luke, Phys. Lett. {\bf B252} (1990) 447; A. F. Falk and B.
Grinstein, Phys. Lett. {\bf B247} (1990) 406; ibid {\bf B249} (1990) 314;
A. F. Falk, B. Grinstein and M. E. Luke, Nucl. Phys. {\bf B357} (1991) 185;
A. F. Falk, H. Georgi and B. Grinstein, Nucl. Phys. {\bf B343} (1990) 1; H.
Georgi, Nucl. Phys. {\bf B348} (1991) 293; H. Georgi, B. Grinstein and M.
B. Wise, Phys. Lett. {\bf B252} (1990) 456.

\item{4.} H. Georgi, Phys. Lett. {\bf B240} (1990) 447.

\item{5.} B. Grinstein, HUTP-91/A028 (1991), HUTP-91/A040 (1991) and
SSCL-17 (1991).

\item{6.} F. Hussain, Proceedings of the ICTP Summer School (1991).

\item{7.} J. Soto and R. Tzani, Phys. Lett. {B297} (1992) 358.
\item{  } J. Soto and R. Tzani, Barcelona preprint UB-ECM-PF-93/1.

\item{8.} L. L. Foldy and S. A. Wouthuysen, Phys. Rev. {\bf 78} (1950)
 29.

\item{9.} F. Feinberg, Phys. Rev. {\bf D17} (1978) 2659.

\item{10.} J. G. K\"orner and G. Thompson, Phys. Lett. {\bf B264}
(1991) 185.

\item{11.} See for example, ``Advanced Field Theory, Micro Macro and
Thermal Physics", H. Umezawa, American Inst. Phys., New York 1993.

\item{12.} See for example, ``Quantum Field Theory", C. Itzykson and J. B.
 Zuber, McGraw Hill (1980); ``Field Theory: A Path Integral Approach",
A. Das, World Scientific (1993).
\end